\documentclass[sigconf]{acmart}

\AtBeginDocument{%
  \providecommand\BibTeX{{%
    \normalfont B\kern-0.5em{\scshape i\kern-0.25em b}\kern-0.8em\TeX}}}

\copyrightyear{2023} 
\acmYear{2023} 
\setcopyright{rightsretained} 
\acmConference[WWW '23]{Proceedings of the ACM Web Conference 2023}{April 30-May 4, 2023}{Austin, TX, USA}
\acmBooktitle{Proceedings of the ACM Web Conference 2023 (WWW '23), April 30-May 4, 2023, Austin, TX, USA}
\acmDOI{10.1145/3543507.3583868}
\acmISBN{978-1-4503-9416-1/23/04}

\PassOptionsToPackage{usenames,dvipsnames}{xcolor}
\usepackage{dblfloatfix}
\usepackage{algorithm}
\usepackage{graphicx}
\usepackage{subcaption}
\usepackage[belowskip=1pt,aboveskip=1pt]{caption}
\usepackage{textcomp}
\usepackage{url}
\usepackage{booktabs}
\usepackage{epstopdf}
\usepackage{multirow}
\usepackage{array}
\usepackage{enumitem}
\usepackage{wrapfig}
\usepackage[many]{tcolorbox}
\usepackage{mathtools}
\usepackage{dsfont}
\usepackage{esvect}
\usepackage{hyperref}
\usepackage{multicol}
\usepackage{xpatch}
\usepackage[noend]{algpseudocode}
\usepackage{comment}
\usepackage{float}
\usepackage{todonotes}
\usepackage{xspace}
\usepackage{booktabs}
\usepackage{arydshln}

\theoremstyle{definition}
\newtheorem{definition}{Definition}[section]

\makeatletter
\xpatchcmd{\algorithmic}{\itemsep\z@}{\itemsep=0.1ex plus0.5pt}{}{}
\makeatother

\begin{document}
\title[Attacking Fake News Detectors via Manipulating News Social Engagement]{Attacking Fake News Detectors via Manipulating\\News Social Engagement}

\author{Haoran Wang$^{1}$, 
        Yingtong Dou$^{2,4}$, 
        Canyu Chen$^{1}$, 
        Lichao Sun$^{3}$, 
        Philip S. Yu$^{2}$, 
        Kai Shu$^{1}$}
\affiliation{
\institution{
$^1$Department of Computer Science, Illinois Institute of Technology, Chicago, IL, USA\\
$^2$Department of Computer Science, University of Illinois Chicago, Chicago, IL, USA\\
$^3$Department of Computer Science and Engineering, Lehigh University, Bethlehem, PA, USA\\
$^4$Visa Research, Palo Alto, CA, USA\\
}
\country{}
}
\email{{hwang219, cchen151}@hawk.iit.edu, 
        {ydou5, psyu}@uic.edu, 
        james.lichao.sun@gmail.com, 
        kshu@iit.edu}

\begin{abstract}
Social media is one of the main sources for news consumption, especially among the younger generation.
With the increasing popularity of news consumption on various social media platforms, there has been a surge of misinformation which includes false information or unfounded claims.
As various text- and social context-based fake news detectors are proposed to detect misinformation on social media, recent works start to focus on the vulnerabilities of fake news detectors.
In this paper, we present the first adversarial attack framework against Graph Neural Network (GNN)-based fake news detectors to probe their robustness.
Specifically, we leverage a multi-agent reinforcement learning (MARL) framework to simulate the adversarial behavior of fraudsters on social media.
Research has shown that in real-world settings, fraudsters coordinate with each other to share different news in order to evade the detection of fake news detectors. 
Therefore, we modeled our MARL framework as a Markov Game with bot, cyborg, and crowd worker agents, which have their own distinctive cost, budget, and influence. 
We then use deep Q-learning to search for the optimal policy that maximizes the rewards.
Extensive experimental results on two real-world fake news propagation datasets demonstrate that our proposed framework can effectively sabotage the GNN-based fake news detector performance.
We hope this paper can provide insights for future research on fake news detection.
\end{abstract}



\maketitle
\section{Introduction}
With the burgeoning of social media, inaccurate or unfounded information (i.e., \textit{misinformation}) is also circulating on social media, which demotes people's belief in truth and science~\cite{shu2017fake, chen2022combating}.
Unlike traditional news articles distributed by news outlets via their media, social engagements like comments and sharing expedite the spread of misinformation and exaggerate its influence at scale.
Recent research has pointed out that misinformation has been hindering the promotion of vaccines and threatening public health during the COVID-19 global pandemic~\cite{loomba2021measuring}.

To combat massive misinformation on social media, many machine learning based misinformation detectors are proposed~\cite{zhou2020survey}.
Besides the methods utilizing natural language processing techniques to check the news content and its writing style to verify its veracity~\cite{ruchansky2017csi, wang2018eann, kaliyarfakebert}, recent works have begun to leverage news social engagement using graph models for fact-checking~\cite{nguyen2020fang, lu-li-2020-gcan, bian2020rumor, ren2020hgat}.
Compared to the straightforward NLP-based methods, social-engagement-based methods regard engaged users as an integral part of news posts.
Based on the theory and evidence that news consumers have preferences on news content (i.e., the \textit{echo chamber})~\cite{dou2021user, shu2019hierarchical, han2020graph, lu-li-2020-gcan, monti2019fake}, the engagement patterns of misinformation and fact are also different.
Moreover, the prevalent bots and fraudsters engaged with fake news posts also differentiate their engagement patterns from regular ones~\cite{shao2018spread}.

Despite the rapid development of automatic fact-checking, most fake news detectors are static models vulnerable to adversarial attacks.
Similar to many security problems, we must acknowledge that misinformation detection is an armed race between content moderators and malicious actors aiming at manipulating public opinion or gaining money through incited social engagement.
Therefore, it is imperative to enhance the robustness of misinformation detectors.
Though some recent works have investigated the robustness of NLP-based misinformation detectors~\cite{horne2019robust, ali2021all, zhou2019fake, le2020malcom, he2021petgen}, no work has probed the robustness of social-engagement-based misinformation detectors.
\cite{le2020malcom} and \cite{lyu2022interpretable} are two closest works to ours. However, they either do not consider social engagement-based detectors or do not model the diverse fraudster type in the misinformation campaign.

\begin{figure}[!tbp]
    \centering
    \includegraphics[width=0.98\linewidth]{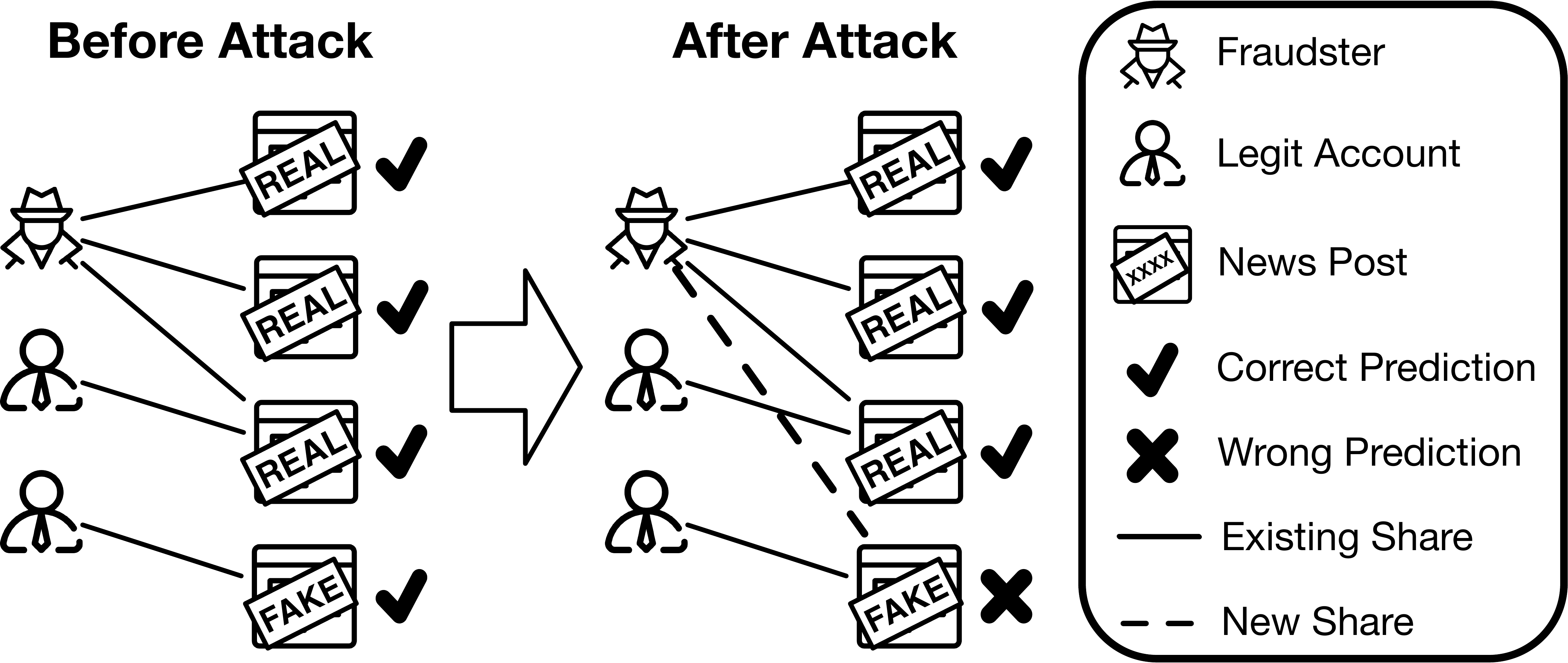}
    \vspace{0.25cm}
    \caption{An illustration of attacking a fake news classifier via manipulating news posts' social engagement.
    The classifier misclassifies the fake news after the fraudster who has shared many real news posts shares it.}
    \Description[Attack toy example]{A toy example of attacking a fake news classifier via manipulating news posts' social engagement. The classifier misclassifies the fake news after the fraudster who has shared many real news posts shares it.}
    \label{fig:demo}
    \vspace{-0.5cm}
\end{figure}

We use Figure~\ref{fig:demo} to demonstrate the vulnerability of social engagement based misinformation detectors. Many existing works~\cite{nguyen2020fang, ren2020hgat} model news social engagement on social media as a heterogeneous graph where users and news posts are nodes, and an edge means a user has shared the post.  
Graph Neural Networks (GNNs)~\cite{kipf2016semi, velivckovic2017graph, hamilton2017inductive} have been widely leveraged to encode the above social engagement graph and predict the veracity of news posts.
Many GNNs are designed to encode the neighboring node information to enhance the prediction performance of the target node.
To exploit this property, as shown in Figure~\ref{fig:demo}, the fraudster who has shared many real news can flip the GNN-based misinformation detector's prediction on a target fake news by sharing it. Because the newly added real news neighbors will alleviate the suspiciousness of the target fake news.

To analyze the robustness of social-engagement-based misinformation detectors,
inspired by GNN robustness research~\cite {sun2018adversarial}, we propose to attack GNN-based misinformation detectors by simulating the adversarial behaviors of fraudsters.
However, the real-world misinformation campaign delivers three non-trivial challenges for attack simulation: 
\textbf{(1)} To evade detection while promoting fake news on social media, malicious actors can only manipulate the controlled user accounts to share different social posts.
However, most of the previous GNN adversarial attack works assume all nodes and edges can be perturbed, which is impractical.
\textbf{(2)} Many deployed GNN-based fake news detectors are grey-box models with various model architectures tailored to the heterogeneous user-post graph.
Thus, the gradient-based optimization method used by previous works~\cite{zugner2018adversarial} cannot be utilized to devise an attack.
\textbf{(3)} Real-world evidence \cite{vargas2020detection, pacheco2020unveiling} shows that various coordinated malicious actors have engaged in the misinformation campaign.
Different types of malicious actors have different capabilities, budgets, and risk appetites.
For instance, key opinion leaders have stronger influence than social bots but cost more to cultivate. 

To overcome the above challenges, we devise a dedicated Multi-agent Reinforcement Learning (MARL) framework, while none of the previous GNN robustness work was used.
Specifically, to simulate the real-world behavior of fraudsters who share different posts, we harness a deep reinforcement learning framework to flip the classification result of a target news node by modifying the connections of users who shared the post. We model the MARL framework as a Markov Game where the agents work coordinately to flip the classification result.
Overall, our contributions are:
\begin{itemize}[leftmargin=*]
    \item To the best of our knowledge, we are the first work to probe the robustness of GNN-based fake news detectors from a social engagement perspective. Although there have been previous works on attacking fake news detectors using NLP methods, attacking fake news detectors by manipulating the social engagement of news targets has not been studied.
    \item We leverage a MARL framework to perform targeted attacks on GNN-based fake news detectors to simulate real-world misinformation campaigns. Specifically, we modeled fraudsters as agents with different costs, budgets, and influences in our framework.
    \item Our experiment results show that our proposed MARL framework could effectively flip the GNN prediction results. We discuss the vulnerabilities of GNN-based fake news detectors and provide insights on attack strategies and countermeasures.
\end{itemize}

The rest of the paper is organized as follows. In Section \ref{sec:related}, we introduce related work. In Section \ref{sec:problem} and \ref{sec03}, we introduce the problem definition and proposed framework. In Section \ref{sec:experiment}, we report our experiment results and analysis. Finally, we discuss the limitation and future work of this paper in Section \ref{sec:conlusion}.
\section{Related Work.}
\label{sec:related}
In this section, we review the related work on (1) graph neural network-based fake news detection; (2) adversarial attack on graph neural networks; and (3) adversarial attack on fake news detection.

\subsection{GNN-based Fake News Detection}
We can categorize the existing GNN-based misinformation detection works into two major categories according to their graph prototypes:
\textbf{1)} Propagation-based work~\cite{lu-li-2020-gcan, monti2019fake, han2021continual, silva2021propagation2vec}: these works model the sharing sequence of a news post as a tree-structured propagation graph with the news post as the root node and edges representing shared relations between users.
It can be formulated as either a propagation graph classification or a root node classification task.
The propagation graph is infeasible for adversarial attacks because the attacker needs to employ a lot of users to share the target post to flip its classification results. At the same time, such operations are naive for optimization and easily captured by simple outlier measurements.
\textbf{2)} Social-context-based work~\cite{nguyen2020fang, ren2020hgat, chandra2020graph, zhou2019network}: all users and their shared news posts (e.g., tweets) formulate a bipartite graph (as shown in Figure~\ref{fig:demo}) where an edge means a user shared the post and the objective is training a GNN to classify the news post nodes.
Note that previous works usually add the publisher as the third type of node connecting to social posts.
In this paper, we only consider the common-used graph prototype (i.e., user-post bipartite graph) as it is easier to manipulate in practice.

\subsection{Adversarial Attack on GNNs}
As GNNs attain excellent performance on many graph mining tasks, their robustness against adversarial attacks has drawn increased attention in recent years~\cite{sun2018adversarial}.
RL-S2V~\cite{dai2018adversarial} and Nettack~\cite{zugner2018adversarial} are two early GNN attacking algorithms aiming at lowering the GNNs' node classification performance via add/deleting edges or modifying node features under a given budget.
Following these work, other works have begun to investigate the GNN robustness under different tasks, e.g., link prediction~\cite{bojchevski2019adversarial}, knowledge graph embedding~\cite{raman2020learning}, and community detection~\cite{li2020adversarial}.
However, none of the previous works have attempted to attack GNN-based fake news detectors, which have recently become popular amid massive adversaries engaging in fake news spread~\cite{shao2018spread}. Compared to the previous works using reinforcement learning to attack GNNs, our work utilizes a multi-agent setting to mimic the real-world misinformation campaign.
In addition, to simulate the real-world attack setting, we only manipulate the edges of the news social engagement graph since it is unlikely that attackers can modify news posts.

\subsection{Adversarial Attack on Fake News Detectors}
Given a wide array of machine learning-based fake news detectors, only a few works have investigated the robustness or vulnerabilities of fake news detectors~\cite{he2021petgen, le2020malcom, zhou2019fake, ali2021all, horne2019robust, koenders2021vulnerable, dou2020robust}.
Among those works, \cite{horne2019robust} examines the robustness of text-based news veracity classifiers over time and against attacks crafted by manipulating news sources.
\cite{ali2021all, koenders2021vulnerable, zhou2019fake} probe the robustness of NLP-based fake news detectors by devising various attacks that distort the news content or inject adversarial texts.
Nash-Detect~\cite{dou2020robust} and AdRumor-RL~\cite{lyu2022interpretable} study the robustness of graph-based spam detectors and rumor detectors respectively, using the reinforcement learning framework.
MALCOM~\cite{le2020malcom} carries out the attacks from another perspective which modifies the comments of each piece of news to fool the fake news detectors leveraging multi-source data.
PETGEN~\cite{he2021petgen} simulates the behavior of malicious users on social media by generating a sequence of texts to attack sequence-based misinformation detectors.
Unlike previous work, we are the first to explore the robustness of social context-based fake news detectors using a multi-agent reinforcement learning framework.
\section{Problem Formulation}
\label{sec:problem}
We formulate the problem of attacking social-engagement-based fake news detectors as attacking GNNs on a user-post sharing graph.
In this section, we first define GNN-based fake news detection and then introduce our adversarial attack objective.

\subsection{GNN-based Fake News Detection}
A user-post sharing graph is defined as a bipartite graph $\mathcal{G}=\{U, V, E, \mathbf{X}_u, \mathbf{X}_v, Y\}$, where $U = (u_0, \cdots, u_i)$ is a set of users, $V = (v_0, \cdots, v_j)$ is a set of news posts, and the edge $e_{ij}=(u_i, v_j) \in E$ indicates the user $u_i$ has shared the news post $v_j$.
$\mathbf{X}_u$ and $\mathbf{X}_v$ are two feature matrices of user nodes and news nodes, respectively.
According to previous works~\cite{dou2021user, nguyen2020fang, han2021continual}, the feature vectors of users and news can be composed of their text representations or handcrafted features.
Following~\cite{dou2021user}, we use the 300-dimensional Glove embeddings of users' historical posts and news post text to represent $\mathbf{X}_u(i, :)$ and $\mathbf{X}_u(j, :)$, respectively.
We use $\mathbf{X}$ to represent all node features for convenience.
$y_j\in Y$ represents the label of $v_j\in V$ where 1 (0 resp.) represents fake news (real news resp.).

To detect fake news based on $\mathcal{G}$, a general GNN framework~\cite{velivckovic2017graph, hamilton2017inductive} can be applied to it.
Concretely, to learn a news post $v$'s representation, a GNN aggregates its neighbors' information recursively:
\begin{equation}\label{eq:gnn_framework}
    \mathbf{h}_{v}^{(l)}= \sigma\left(\mathbf{h}_{v}^{(l-1)} \oplus \text{AGG}^{(l)}\left(\mathbf{h}_{u}^{(l-1)}, (u,v)\in E \right)\right),
\end{equation}
where $l$ is the GNN layer number.
AGG is the GNN aggregator that aggregates neighbor embeddings. Common aggregators employ attention~\cite{velivckovic2017graph}, mean~\cite{hamilton2017inductive}, and summation~\cite{hamilton2017inductive}.
$\oplus$ represents the operation that combines the embedding of $v$ at the last GNN layer and its aggregated neighbor embeddings.
Common approaches include concatenation and summation.
Similarly, the representation of $u$ can be learned by the same process shown in Eq. (\ref{eq:gnn_framework}).

To classify the news post $v\in V$, a GNN classifier $f$ takes the $\mathcal{G}$ as input where $X_u$ and $X_v$ are node features of $u$ and $v$.
It maps $v\in V$ to $y\in(0,1)$ after feeding the $h_v$ at the last layer to an MLP and softmax layer.
The GNN classifier can be trained on partially labeled post nodes with the following cross-entropy loss in a semi-supervised fashion:
\begin{equation}\label{eq:gnn_loss}
\mathcal{L}_{GNN}(\mathcal{G}, f_\theta) =
\sum_{v_j\in \mathcal{V}}-\text{log}\left(y_{j}\cdot\sigma((f_\theta(E, \mathbf{X})_j)\right).
\end{equation}

\subsection{Adversarial Attacks on GNN-based Fake News Detectors}
\label{sec02:attack}
At a high level, our problem can be regarded as attacking GNN-based node classification but with practical constraints to simulate real-world misinformation campaigns. Specifically, the objective of the attacking method is to flip the GNN classification results of target social posts via maneuvering controlled malicious social media user accounts to share new posts. Note that we assume attackers can only perturb the graph by controlling malicious users to share news posts and not delete existing shared news posts. We make this assumption because in a real-world setting, even though the users can delete existing shared posts, the record of shared relations may still exist in the database.
Considering the massive social network data and the diverse fake news detectors employed by the platform, 
we assume the unknown target GNN is pre-trained on clean data in our problem setting (i.e., the training data is not poisoned by the adversary). Also, we assume that we have knowledge about the type of GNN the detector is trained on, but we do not have access to its model parameters.
Thus, our problem is a grey-box evasive structural attack on the GNN-based node classification task.
We formally define our attack objective as:
\begin{equation}\label{eq:attack_obj}
    \begin{aligned}
    & \underset{U_{c}, E_{a}}{\text{max}}
     \sum_{v\in V_T} \mathds{1}(f_{\theta^*}(E^{\prime}, \mathbf{X})_v \neq y_v) \\
    & \text{s.t.} \ \
     {\theta ^*} = \mathop {\arg \min }\limits_\theta \mathcal{L}_{GNN}(\mathcal{G}, f_\theta), \\
    & \qquad |U_{c}| \leq \Delta_u,  |E_{a}| \leq \Delta_e,
    \end{aligned}
    \end{equation}
where $U_{c}$, $E_{a}$, and $V_T$ represent a set of controlled users, manipulated edges, and target news posts, respectively.
$\mathcal{G}$ represents the clean graph and $E^\prime$ is the set of perturbed edges.
$\Delta_{u}$ ($\Delta_{e}$ resp.) represents the budget of controlled users (modified edges resp.).
The above adversarial objective essentially maximizes the misclassification rate of the target social posts.
\section{Methodology}
\label{sec03}
In practice, misinformation campaigns are carried out by coordinated fraudsters manipulating social user accounts to evade detection. In this section, we first elucidate the property of attackers motivated by real-world misinformation campaigns. Then we present the multi-agent reinforcement learning framework we use to probe the robustness of GNN-based fake news detectors.

\subsection{Attacker Property}
\subsubsection{Attacker Knowledge}
As introduced in Section~\ref{sec02:attack}, our attack is a grey-box attack meaning the attackers only have knowledge about the architecture of the GNN-based classifier, but not its model internals like weights or coefficient values.
To attack the GNN classification results of target posts, we assume the attacker can sense the entire graph, including the features and labels of the user and post nodes as well as their connections.
The above setting is practical since social media information is publicly accessible and the attackers can easily infer node features and labels given fruitful related works in misinformation research.

\subsubsection{Attacker Capability}
\label{sec3:capability}
To imitate the real-world behavior of fraudsters as much as possible, we define the capability of users controlled by fraudsters (i.e., $U_c$) as follows:

\begin{itemize}[leftmargin=*]
\item \textit{Direct Attack}: $u\in U_c$ shares the $v\in V_T$ directly if $(u, v)\notin E$.
In real-world settings, given that a controlled user shared many posts from trustworthy sources that seemed to be legitimate, it will help alleviate the suspiciousness of a fake news post if the user shares the post.  

\item \textit{Indirect Attack}: For $u\in U_c, v\in V_T, (u, v)\in E$, we carry out the attack by controlling $u$ to share $v^\prime \notin V_T$.
The indirect attack exploits the neighbor aggregation mechanism of GNNs by exerting influence on the target post by changing its neighborhood.
In practice, for a controlled user having shared the target post with fake news, one can let the controlled user share posts from trustworthy sources to mislead the GNN's prediction on the target fake news post.
\end{itemize}

Note that as previously mentioned in Section \ref{sec02:attack}, attackers are only allowed to add edges between user nodes and news post nodes.

\subsubsection{Agent Configuration.}

\begin{figure}[tbp]
\centering
\vspace{-0.1cm}
\resizebox{0.90\linewidth}{!}{%
    \begin{subfigure}[b]{0.23\textwidth}
    \centering
    \includegraphics[width=\textwidth]{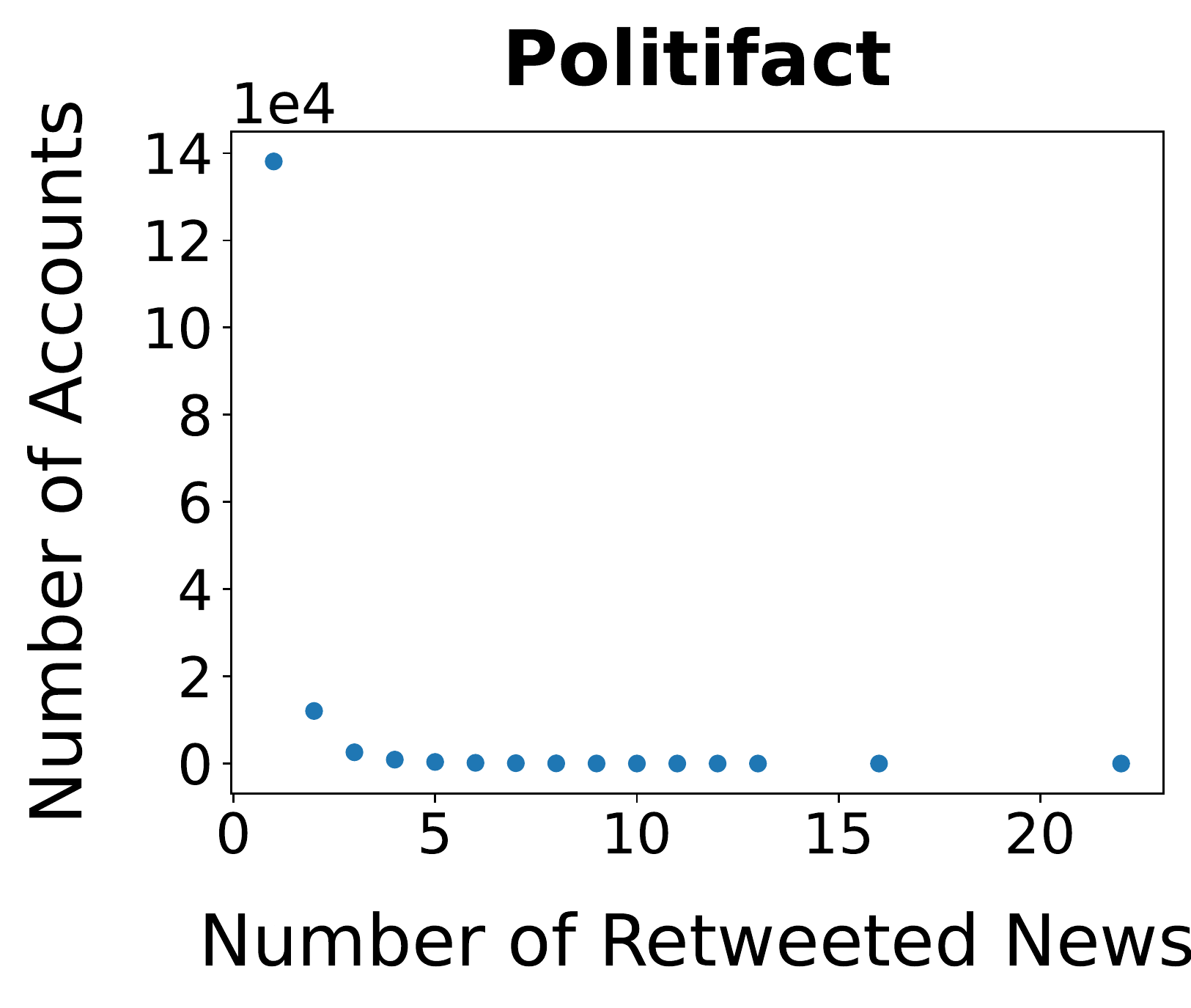}
    \end{subfigure}
    \hspace{0.2cm}
    \begin{subfigure}[b]{0.23\textwidth}
    \centering
    \includegraphics[width=\textwidth]{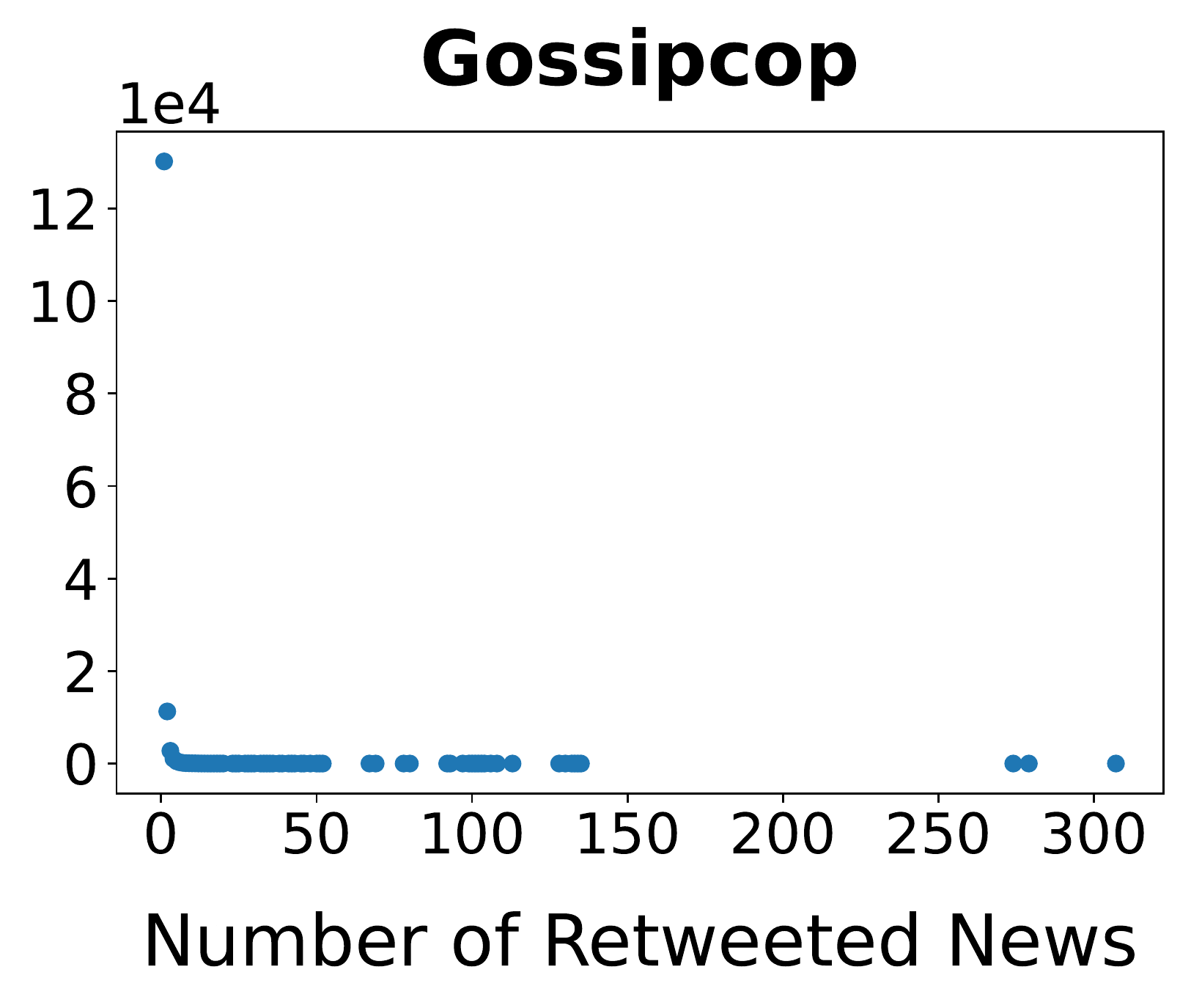}
    \end{subfigure}}
    \vspace{0.25cm}
 \caption{The user account number distribution according to the amount of news shared by them. We sample accounts in different ranges to represent different user types.}
 \Description[User account number distribution]{The user account number distribution according to the amount of news shared by them. We sample accounts in different ranges to represent different user types.}
 \label{fig:distribution}
 \vspace{-0.25cm}
\end{figure}

\label{sec03:agent}
Real-world evidence shows that multiple groups of malicious actors are engaged in the misinformation campaign~\cite{shao2018spread, pacheco2021uncovering, shu2017fake}.
Besides singletons who act individually, most misinformation campaigns are executed coordinately by professional agencies since it would reach the campaign goal faster while maximizing the utilities of existing resources.
From the adversarial attack perspective, different types of controlled user accounts have distinct influences on target posts and different budgets.
For instance, the bot users are usually low-cost and with a higher budget.
However, these bot users have few historical records; thus each bot user has limited influence on target posts~\cite{shao2018spread}.
The crowd workers with credible and rich social profiles are usually expensive, but they have a stronger influence on target posts.

\begin{table}[!b]
    \centering
    \captionsetup{width=0.98\linewidth}
    \caption{The comparison of properties among bots, cyborgs, and crowd worker agents.}
    \vspace{0.1cm}
    \resizebox{0.95\linewidth}{!}{%
        \begin{tabular}{@{}ccccc@{}}
        \toprule
        \textbf{Agent} & \textbf{User}      & \textbf{Cost}     & \textbf{Influence} & \textbf{Budget}   \\ \midrule
        \textbf{1}     & bot          & low      & low       & high     \\
        \textbf{2}     & cyborg       & moderate & moderate  & moderate \\
        \textbf{3}     & crowd worker & high     & high      & low      \\ \bottomrule
        \end{tabular}}
    \label{tab:agent}
\end{table}

To model the above distinct malicious actor groups, previous single-agent RL frameworks are not applicable~\cite{dai2018adversarial, sun2020non}.
Therefore, we leverage a MARL, which not only enables the personalized configuration for each group but also helps simulate the coordinated behavior between different groups.
Specifically, we define three agents which control three distinct groups of user accounts according to the malicious accounts introduced in~\cite{shu2017fake}.
We divide the user accounts based on the number of news they have shared, Figure \ref{fig:distribution} shows the distribution of the number of news that users have shared in Politifact and Gossipcop datasets.
Table~\ref{tab:agent} compares the key properties of the following agents.

\noindent 1) \textbf{Agent 1 (Social Bots):}
Social bots registered and fully controlled by automated programs have been proven to engage in fake news spreading by many works~\cite{bessi2016social, shao2018spread}.
The first agent controls the bot users, and it has a low cost and high budget.
We randomly select the users with only one connection in our datasets to represent the newly created bot users.

\noindent 2) \textbf{Agent 2 (Cyborg Users):}
According to \cite{shu2017fake}, cyborg users are
registered by humans and partially controlled by automated programs.
The easy switch of functionalities between humans and bots offers cyborgs unique opportunities to spread fake news.
Since those users are camouflaged as human, they usually have more historical engagements (i.e., connections to other posts).
In our datasets, we randomly select the users with more than 10 connections to represent the compromised users.
The cost, budget, and influence of cyborg agents are between that of the other two agents.

\noindent 3) \textbf{Agent 3 (Crowd Workers):}
The crowd workers are usually of high cost since they get paid for each campaign.
Meanwhile, they have the strongest influence.
We take the users with more than 20 connections, where 100\% of them connect to real-news posts to represent the crowd workers.

\subsection{Attack Framework}
\label{sec03:framework}

\begin{figure}[tbp]
    \centering
    \includegraphics[width=0.98\linewidth]{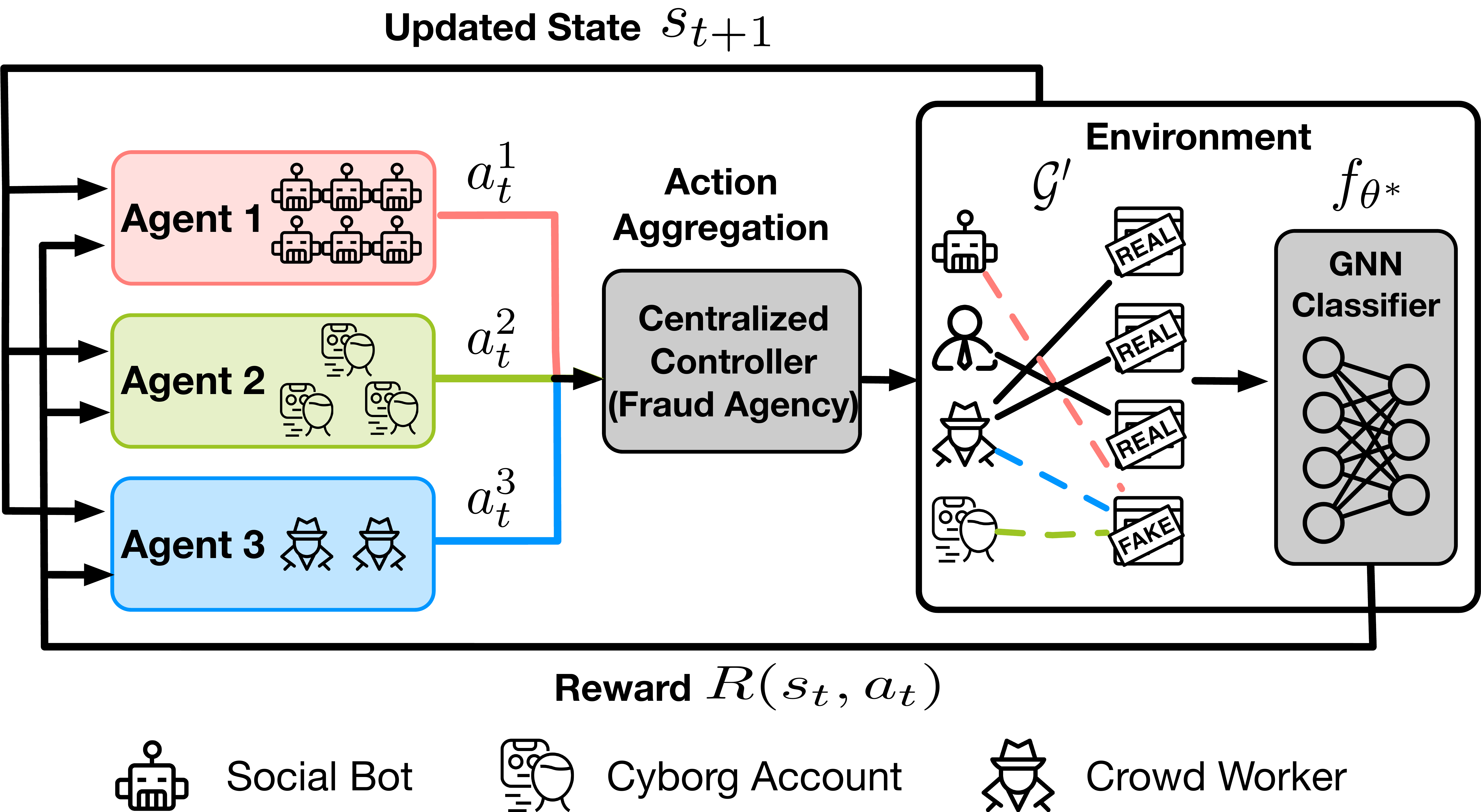}
    \vspace{0.2cm}
    \caption{The proposed MARL framework to generate adversarial edge perturbations against GNN-based fake news classifier. See Section~\ref{sec03:framework} for more details.}
    \Description[Attack framework]{The proposed MARL framework to generate adversarial edge perturbations against GNN-based fake news classifier.}
    \label{fig:framework}
    \vspace{-0.5cm}
\end{figure}

In the real world, each agent above represents a malicious actor that aims to influence the fake news classification results.
Given a set of target news posts $V_T$, the attack process can be modeled as a multi-agent cooperative reinforcement learning problem where all agents work together to maximize the misclassification rates of target news posts.
Figure~\ref{fig:framework} illustrates the attack process of the proposed MARL algorithm.
First, actions made from different agents are aggregated by the center controller;
then, aggregated actions are applied to the environment composed of the social engagement graph and the surrogate classifier;
the updated state and rewards generated by the classifier are finally sent back to each agent for the next episode of optimization.
In this subsection, We first define each component of the MARL framework, then introduce how we leverage deep Q-learning for optimization.

\subsubsection{MARL Framework}

Different from previous GNN attacks using single-agent RL, which can be modeled as a Markov Decision Process (MDP)~\cite{dai2018adversarial, sun2020non},
the MARL framework is a Markov game (MG).
We formally define the MG and its components as follows:

\definition
A Markov game is defined by a tuple \\
$(\mathcal{N}, \mathcal{S}, \{\mathcal{A}^{i}\}_{i\in\mathcal{N}}, \mathcal{P}, \{R^{i}\}_{i\in\mathcal{N}}, \gamma)$, where $\mathcal{N} = \{1, \cdots, N\}$ represent the set of $N$ agents, $\mathcal{S}$ is the state space observed by all agents, $\mathcal{A}^{i}$ denotes the action space of agent $i$.
$\mathcal{P}$ is the state transition probability given a state $s\in \mathcal{S}$ and action $a\in \mathcal{A}$.
$R$ is the reward function that determines the immediate reward received by agent $i$ after a transition from $(s, a)$ to $s^\prime$.
$\gamma$ is the discount factor to long-term reward.

\begin{itemize}[leftmargin=*]
    \item \textbf{Action.}
    As defined in Section~\ref{sec3:capability}, each $u\in U_c$ can only add edges based on their connection status to $v\in V_T$.
    Meanwhile, each agent controls a set of users $ U_{c}^{i}$ according to Section~\ref{sec03:agent}.
    We use $a^{i}_{t}(u, v)$ to denote the action that adds the edge between user $u$ and post $v$.
    Thus, the action space for agent $i$ at time $t$ is $a^i_t \in \mathcal{A}^i\subseteq U_{c}^{i}\times V_T $.
    We use a centralized controller to aggregate agent actions. Specifically, in each episode, the final actions are from the three types of agents with a fixed proportion, this proportion is motivated by real-world misinformation campaigns.

    \item \textbf{State.}
    Since all agents work cooperatively to attack the same set of target posts $V_T$ against the same classifier $f$, all agents share the same state at time $t$ represented by $(\mathcal{G}^\prime_{t}, f)$, where $\mathcal{G}^\prime_{t}$ is the perturbed graph at time $t$.

    \item \textbf{Reward.}
    As a grey-box attack, we aim to flip the classification results of the target classifier.
    Since we have knowledge of the GNN architecture of the detector, we use one of the three GNN models (i.e. GAT, GCN, and GraphSAGE) as our surrogate target classifier and take its classification results on $V_T$ as the reward to guide the agent.
    Note that the reward is shared by all agents under the cooperative setting.
    After all agents make the actions under their budgets (i.e., one episode), the reward for each agent towards the target post $v\in V_T$ is:
    \begin{equation}
    R\left((\mathcal{G}^\prime, f_{\theta^*})_v\right) = \left\{\begin{array}{l}1: f_{\theta^*}(E^{\prime}, \mathbf{X})_v \neq y_v,  \\ -1: f_{\theta^*}(E^{\prime}, \mathbf{X})_v = y_v. \end{array}\right.
    \label{eq:reward_func}
    \end{equation}
    
    \item \textbf{Terminal.}
    After each agent makes finite modifications according to their own budget $\Delta^i_e$, the Markov game stops.
\end{itemize}

\subsubsection{Deep Q-Learning.}
To solve the above Markov game, we need to find the optimal policy that maximizes the expected value of long-term rewards.
Since each agent has its own controlled user accounts and budget, each agent $i$ should have its own policy $\pi^i$ that $a^i_t \sim \pi^{i}(\cdot|s_t)$.
We use the Q-learning to learn the optimal policy $\pi^{i, \ast}$ parameterized by a Q-function $Q^{i, *}(s_t, a_t)$.
The optimal Q-value for agent $i$ can be represented by the following Bellman equation:
\begin{equation}
    Q^{i, *}\left(s_{t}, a^{i}_{t}\right)=R\left(s_{t}, a^{i}_{t}\right)+\gamma \max _{a^{\prime}} Q^{*}\left(s_{t+1}, a^{i, \prime}\right),
    \label{eq:q_function}
\end{equation}
where $a^{i, \prime}$ represents agent $i$'s future action based on state $s_t$.
The above equation suggests a greedy policy where the agent $i$'s best action based on $s_t$ is the action that maximizes the Q-value above:
\begin{equation}
    \pi^{i}\left(a^{i}_{t} \mid s_{t} ; Q^{i, *}\right)=\mathop {\arg \max }\limits_{a^{i}_{t}} Q^{i, *}\left(s_{t}, a^{i}_{t}\right).
    \label{eq:q_greedy}
\end{equation}

For each target post $v\in V_T$, we would like to choose the controlled user $u\in U_c$ with the most influence to $v$ to flip the GNN classification result on $v$.
Thus, using GNNs to parameterize the Q-function could help model each action's value. 
Specifically, we first employ a two-layer GraphSAGE~\cite{hamilton2017inductive} to obtain the embedding of each post node $h_{v,t}$ in current state $s_t$ according to Eq.~(\ref{eq:gnn_framework}).
Note that we only consider the 2-hop neighborhood of all target nodes and controlled user accounts, which could reduce the state and action space.
For agent $i$ at time $t$ with embeddings of controlled user accounts $h_{u, t}, u\in U^{i}_{c}$ and the target node $h_{v,t}, v\in V_{T}$, the Q-value of action $a^{i}_{t}(u , v)$ is calculated by the following equation:
\begin{equation}
     Q^{i}\left(s_{t}, a^{i}_{t}(u, v)\right) = \sigma\left(W^{1}(h_{u, t})\right) \cdot  \sigma\left(W^{2}(h_{v, t})\right),
    \label{eq:q_parameter}
\end{equation}
where two liner layers are applied on two end node embeddings before computing their dot product which yields the Q-value of the given action.

\begin{table}[!tbp]
  \centering
  \captionsetup{width=0.98\linewidth}
    \caption{Dataset statistics and agent configurations for Politifact and Gossipcop datasets.}
    \vspace{0.1cm}
  \resizebox{0.98\linewidth}{!}{%
    \begin{tabular}{cccccccc}
    
    \hline
    \textbf{Data}                 & \textbf{$|U|$}           & \textbf{$|V|$}          & \textbf{$|E|$}             & \textbf{$|V_T|$}       & \textbf{Agent} & \textbf{$\Delta_u$} & \textbf{$\Delta_e$} \\ \hline
    \multirow{3}{*}{\textbf{POL}}     & \multirow{3}{*}{276,277} & \multirow{3}{*}{581}    & \multirow{3}{*}{1,074,890} & \multirow{3}{*}{62}    & 1 & 100    & 1 \\
         &   &   &   &    & 2   & 50  & 3   \\
         &   &   &   &    & 3   & 20  & 5   \\ \hline
    \multirow{3}{*}{\textbf{GOS}}     & \multirow{3}{*}{565,660} & \multirow{3}{*}{10,333} & \multirow{3}{*}{3,084,931} & \multirow{3}{*}{1,547} & 1 & 1,000 & 1 \\
         &   &   &   &    & 2   & 500 & 3   \\
         &   &   &   &    & 3   & 100 & 5   \\ \hline
    \end{tabular}
  }
  \label{tab:dataset}
\end{table}

\begin{table}[!b]
\centering
\captionsetup{width=0.80\linewidth}
\caption{Performance of surrogate models measured by accuracy and F-1 score.}
\vspace{0.1cm}
\resizebox{0.80\linewidth}{!}{%
\begin{tabular}{@{}ccccc@{}}
\toprule
\multirow{2}{*}{\textbf{Models}} & \multicolumn{2}{c}{\textbf{Politifact}} & \multicolumn{2}{c}{\textbf{Gossipcop}} \\ 
\cmidrule(l){2-3} \cmidrule(l){4-5}
                        & Accuracy        & F1           & Accuracy       & F1           \\ \midrule
\textbf{GCN}                     & 0.8673          & 0.8632       & 0.8278         & 0.7864       \\
\textbf{GAT}                     & 0.8600          & 0.8543       & 0.8423         & 0.8010       \\
\textbf{SAGE}               & 0.8034          & 0.7973       & 0.8824         & 0.8636       \\ \bottomrule
\end{tabular}}
\label{tab:model}
\end{table}

\begin{table*}[!tbp]
  \centering
  \captionsetup{width=0.75\linewidth}
    \caption{Results of using MARL to perform \textit{indirect} targeted attacks comparing to several baselines. Experiments are repeated five times, and the average success rate is reported.}
    \vspace{0.1cm}
  \resizebox{0.75\linewidth}{!}{%
    \begin{tabular}{@{}ccccccccccccc@{}}
      \toprule    
      \multirow{3}{*}{\textbf{Method}} & \multicolumn{6}{c}{\textbf{Politifact}}                             & \multicolumn{6}{c}{\textbf{Gossipcop}}                             \\ 
      \cmidrule(lr){2-7} \cmidrule(lr){8-13}
      & \multicolumn{3}{c}{\textbf{Fake}} & \multicolumn{3}{c}{\textbf{Real}} & \multicolumn{3}{c}{\textbf{Fake}} & \multicolumn{3}{c}{\textbf{Real}} \\ 
      \cmidrule(lr){2-4} \cmidrule(lr){5-7} \cmidrule(lr){8-10} \cmidrule(lr){11-13}
                             & GAT           & GCN           & SAGE          & GAT           & GCN           & SAGE          & GAT           & GCN           & SAGE          & GAT           & GCN           & SAGE          \\ \midrule
      \textbf{RD-Edge}   & 0.14          & 0.45\         & 0.13          & 0.11          & 0.33          & 0.15          & 0.06          & 0.28          & 0.25          & 0.08          & 0.22          & 0.14          \\
      \textbf{RD-Node}   & 0.12          & 0.48          & 0.14          & 0.13          & \textbf{0.38} & 0.15          & 0.12          & 0.32          & 0.22          & 0.12          & 0.23          & 0.16          \\
      \textbf{RL - A1}       & 0.17          & 0.82          & 0.16          & 0.08          & 0.07          & \textbf{0.21}          & 0.14          & 0.45          & 0.23          & 0.08          & 0.80          & 0.16          \\
      \textbf{RL - A2}       & 0.15          & 0.91          & 0.18          & 0.08          & 0.13          & 0.18          & 0.18          & 0.52          & 0.32          & 0.06          & 0.83          & 0.24          \\
      \textbf{RL - A3}       & 0.08          & 0.86          & 0.13          & 0.08          & 0.13          & 0.18 & 0.16          & 0.42          & 0.28         & 0.12          & 0.85          & 0.22          \\ \hdashline
      \textbf{MARL} & \textbf{0.33} & \textbf{0.92} & \textbf{0.28} & \textbf{0.22} & 0.31          & 0.19          & \textbf{0.21} & \textbf{0.64} & \textbf{0.36} & \textbf{0.18} & \textbf{0.89} & \textbf{0.28} \\ \bottomrule
    \end{tabular}
}
  \vspace{-0.25cm}
  \label{tab:performance}
\end{table*}

\section{Experiments} \label{sec:experiment}
As mentioned in Section \ref{sec3:capability}, indirect attacks do not modify the edges between user nodes and target news nodes directly and are more likely to be used by attackers in practice to evade detection. In this section, we conduct a series of experiments to validate the effectiveness of our proposed framework under the more realistic \textbf{indirect} settings and analyze the factors that affect MARL's attack performance. Then, we present experiment results to compare direct attack against indirect attack. Finally, we discuss some countermeasures that could be used by defenders.
Specifically, we aim to answer the following research questions:

\begin{itemize}[leftmargin=*]
\item \textbf{RQ1}: How does the performance of MARL compare to baselines?
\item \textbf{RQ2}: What factors affect the performance of MARL?
\item \textbf{RQ3}: How does direct attack compare with indirect attack?
\item \textbf{RQ4}: What are some countermeasures against attacks?
\end{itemize}

\subsection{Experiment Settings}
In this subsection, we introduce the experiment settings for MARL indirect attacks. We first introduce the datasets, surrogate models, and baseline methods used for the experiments, then we introduce the implementation details.

\subsubsection{Datasets}
We extract two social engagement graphs from the FakeNewsNet~\cite{shu2018fakenewsnet} dataset composed of the metadata of fake and real news posts and their engaged users on Twitter from two fact-checking sources: Politifact and Gossipcop.
Following~\cite{dou2021user}, we take Glove 300D \cite{pennington-etal-2014-glove} embedding of a user's historical tweets as its feature and the Glove embedding of the associated news content of a social post as its features.
Note that our attack operates the controlled users to share posts; since the number of changed edges for a user is within a tight budget, we assume all node features are unchanged during the attack process.

\subsubsection{Surrogate Models}
Under the grey-box setting, the attacker only has information about the architecture of the model being attacked. Thus the attack has to be performed on a surrogate model $\mathcal{M}$ that has the same GNN architecture as the target model. 
For the GNN-based fake news detectors, we include three classic GNN models. Specifically, we use Graph Convolution Network \cite{kipf2016semi}, Graph Attention Network \cite{velivckovic2017graph}, and GraphSAGE\cite{hamilton2017inductive} as our $\mathcal{M}$. Table \ref{tab:model} shows the performance of these surrogate models on Politifact and Gossipcop. We trained these models to ensure that they have similar performance across both datasets. So that we can measure the attack performance of MARL comparably.

\subsubsection{Baseline Attack Methods}
Due to the attacker's limited capability and restricted candidates of both controlled users and target posts, we cannot take the feature and gradient-based attacks~\cite{zugner2018adversarial} as baselines.
To compare the effectiveness of the proposed MARL framework, we compare it with the following baselines:
\begin{itemize}[leftmargin=*]
  \item Random-Edge (RD-Edge):
        This is a simple baseline that randomly selects the controlled users and target posts to add edges until meeting the budget.
  \item Random-Node (RD-Node):
        This baseline injects new user nodes into the graph and connects them with the target news nodes.
  \item Single Agent RL:
        To demonstrate the effectiveness of MARL, we created this baseline by limiting the attacker to a single type of agent. Specifically, we have three baselines named RL-A1 (Bot), RL-A2 (Cyborg), and RL-A3 (Crowd Worker).
\end{itemize}

\noindent \textbf{Budget and Target Selection Criteria} For the Politifact dataset, we randomly sampled 100 bot agents, 50 cyborg agents, and 20 crowd worker agents from Table 1. For the Gossipcop dataset, we randomly sampled 1,000 bot agents, 500 cyborg agents, and 100 crowd worker agents.

\noindent \textbf{Implementation Details} 
For Random-Edge method, we connect edges between sampled attack agents and news targets based on the agent node's degree. Specifically, we randomly connect bot agents with 1 news target, cyborg agents with 3 news targets, and crowd worker agents with 5 new targets.  For Random-Node method, we add 5 user nodes for each of the three agent types. We generate the embedding for each node by randomly sampling 20 nodes from each type of agent, and taking the average of their embedding as the new embedding for the injected node. We connect the generated user nodes with target news nodes the same way in the Random-Edge method.
We use PyG~\cite{Fey2019fast} to implement all GNN algorithms.
The MARL algorithm is implemented based on the RL-S2V code provided by~\cite{dai2018adversarial}.
Our code and data are publicly available \footnote{\url{https://github.com/hwang219/AttackFakeNews}}.

\noindent \textbf{Performance Metrics} Since we only aim to flip the classification results of a selected group of target posts, we use the success rate (SR) as the metric to evaluate the attack performance, which is the number of misclassified posts divided by the total number of target posts after the attack.

\begin{figure*}[tbp]
  \centering
\resizebox{0.95\linewidth}{!}{%
  \begin{subfigure}[b]{0.32\textwidth}
    \centering
    \includegraphics[width=\textwidth]{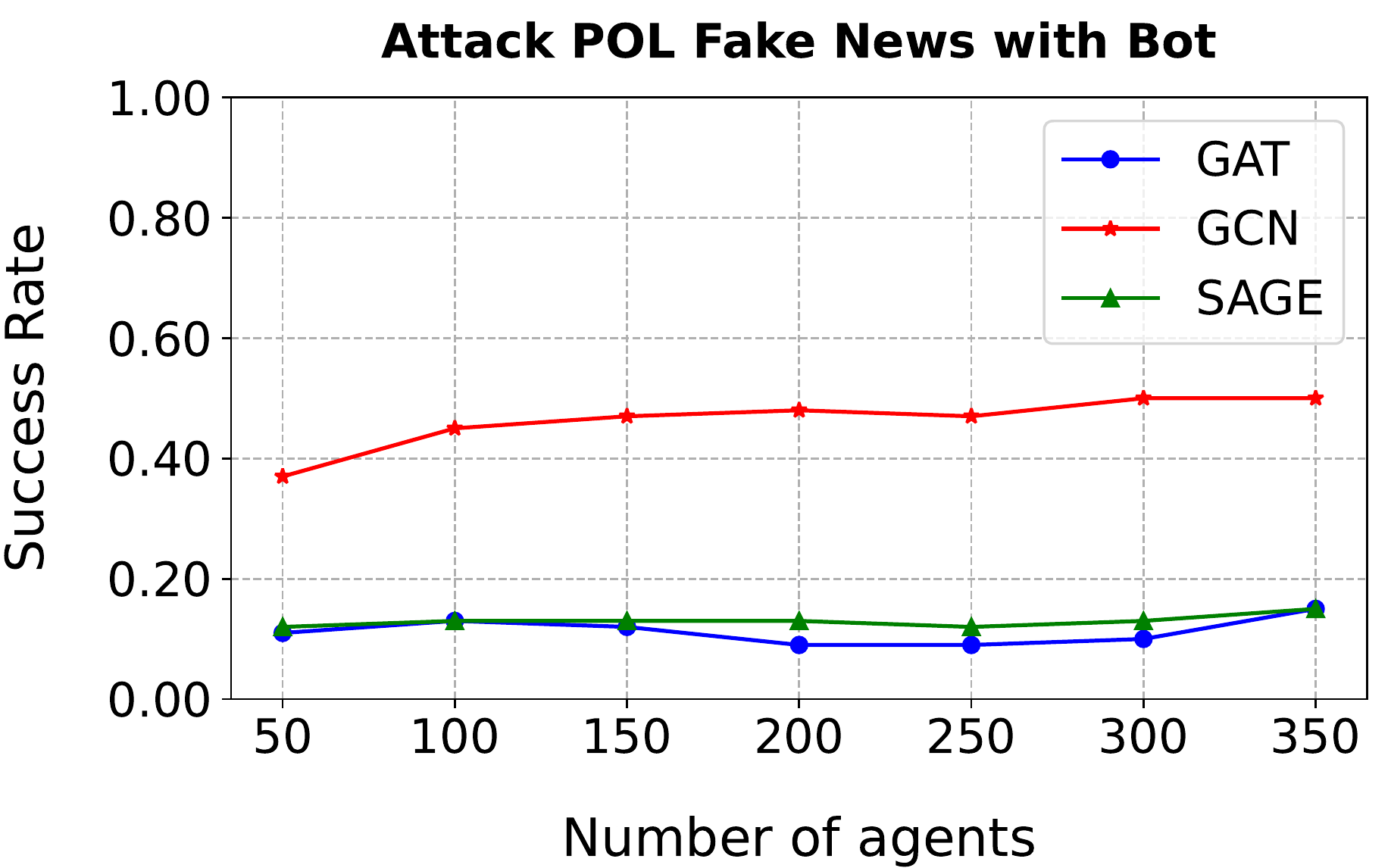}
  \end{subfigure}
  \hfill
  \begin{subfigure}[b]{0.32\textwidth}
    \centering
    \includegraphics[width=\textwidth]{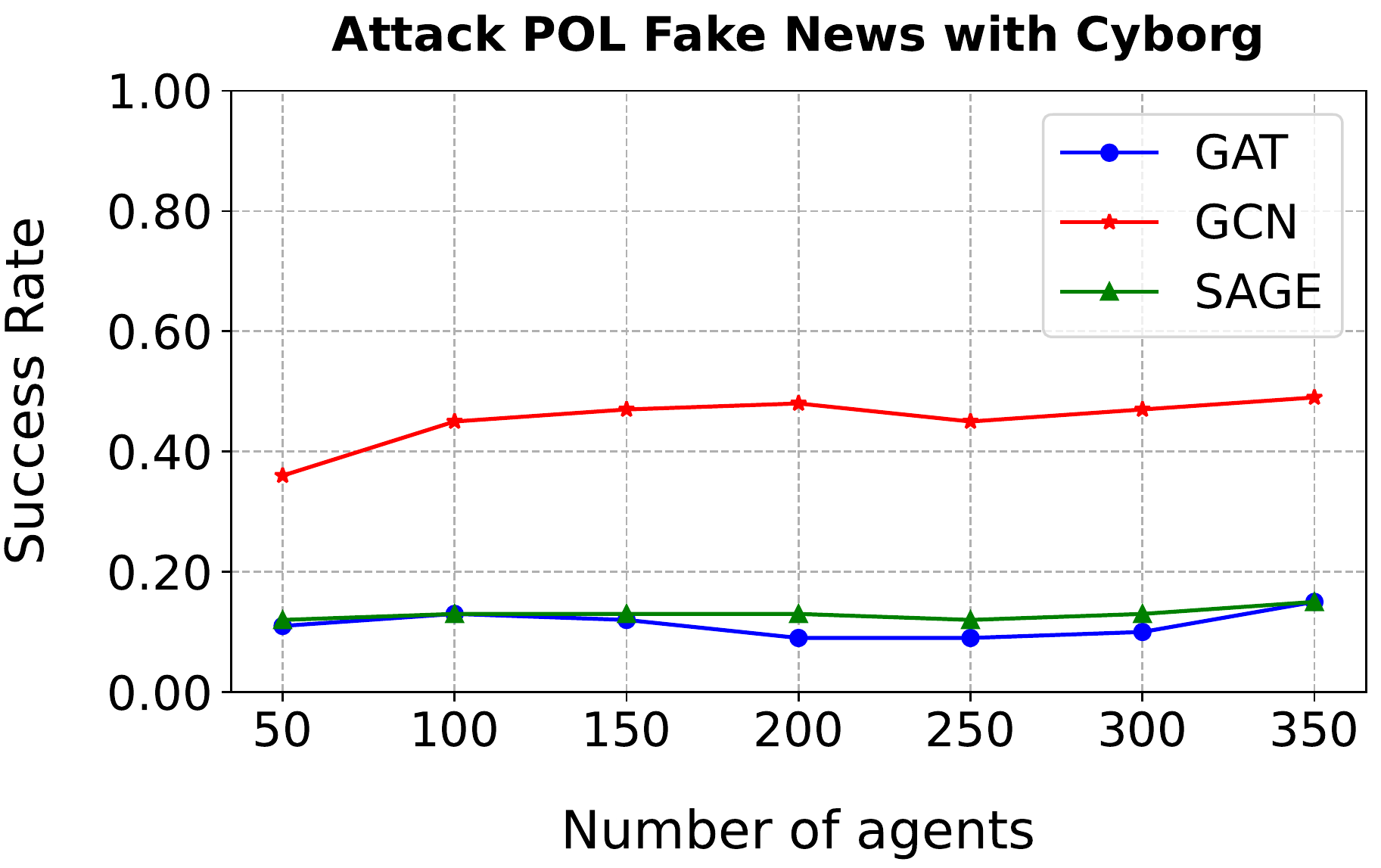}
  \end{subfigure}
  \hfill
  \begin{subfigure}[b]{0.32\textwidth}
    \centering
    \includegraphics[width=\textwidth]{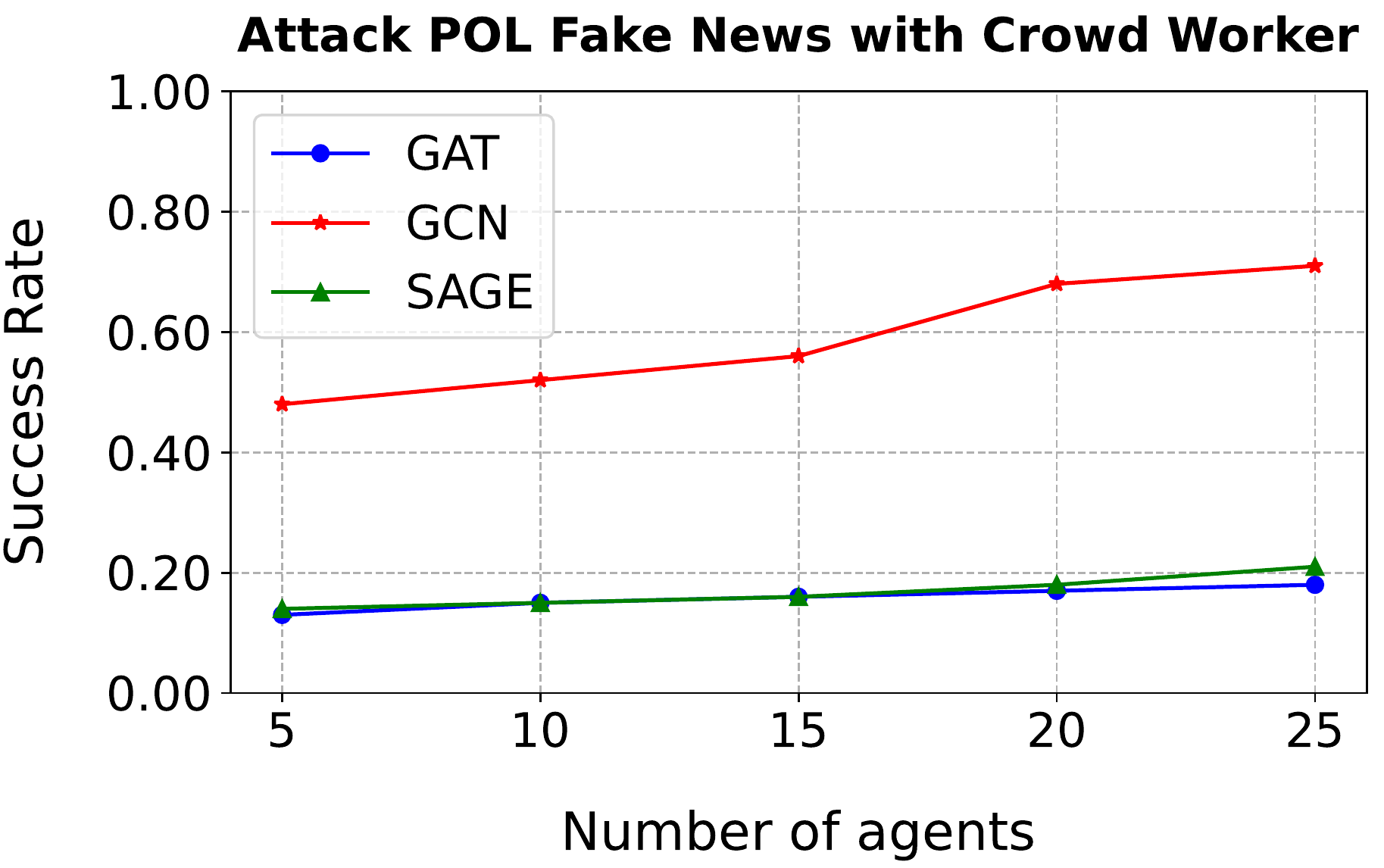}
  \end{subfigure}}
\resizebox{0.95\linewidth}{!}{%
  \begin{subfigure}[b]{0.32\textwidth}
    \centering
    \includegraphics[width=\textwidth]{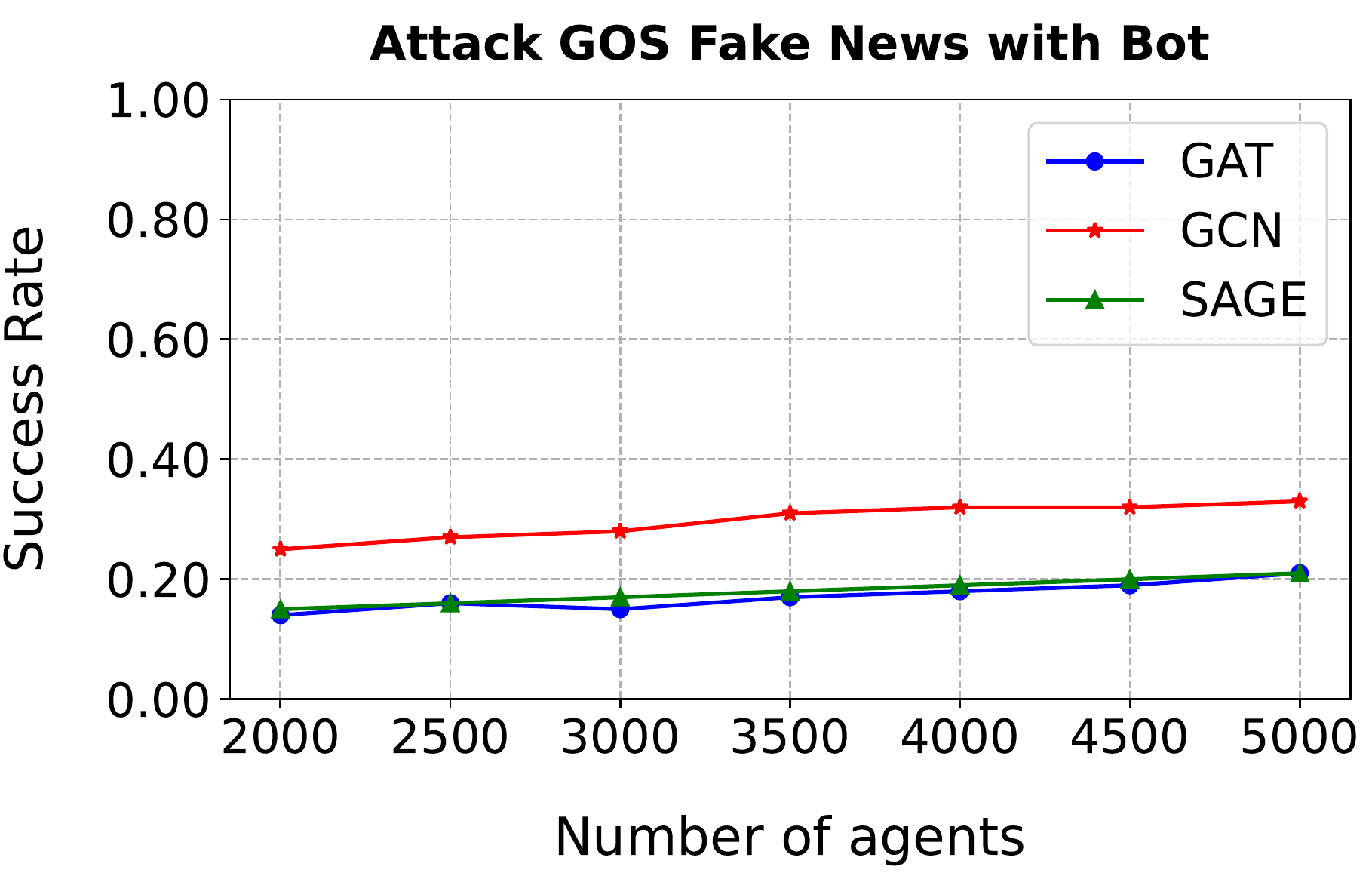}
  \end{subfigure}
  \hfill
  \begin{subfigure}[b]{0.32\textwidth}
    \centering
    \includegraphics[width=\textwidth]{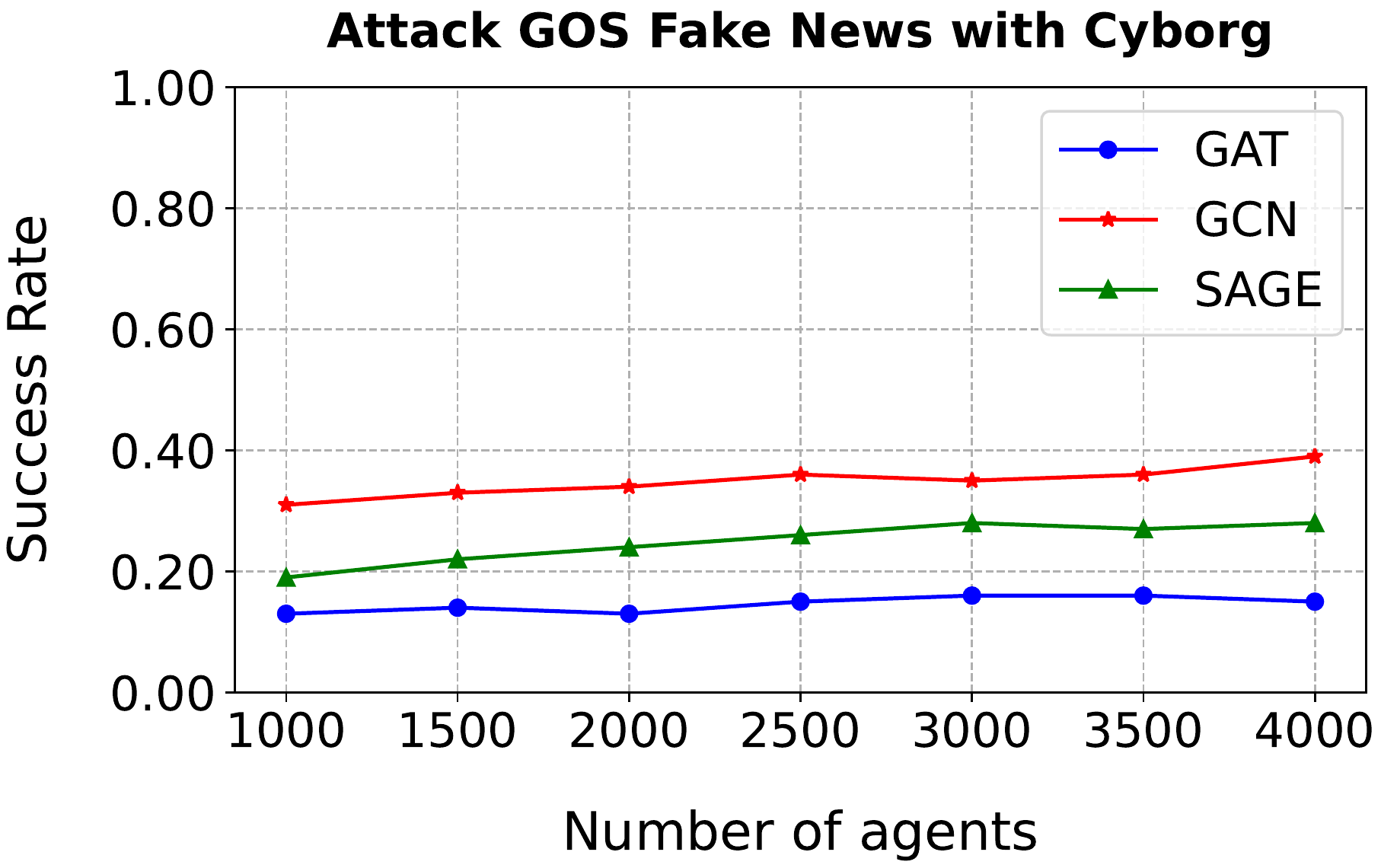}
  \end{subfigure}
  \hfill
  \begin{subfigure}[b]{0.32\textwidth}
    \centering
    \includegraphics[width=\textwidth]{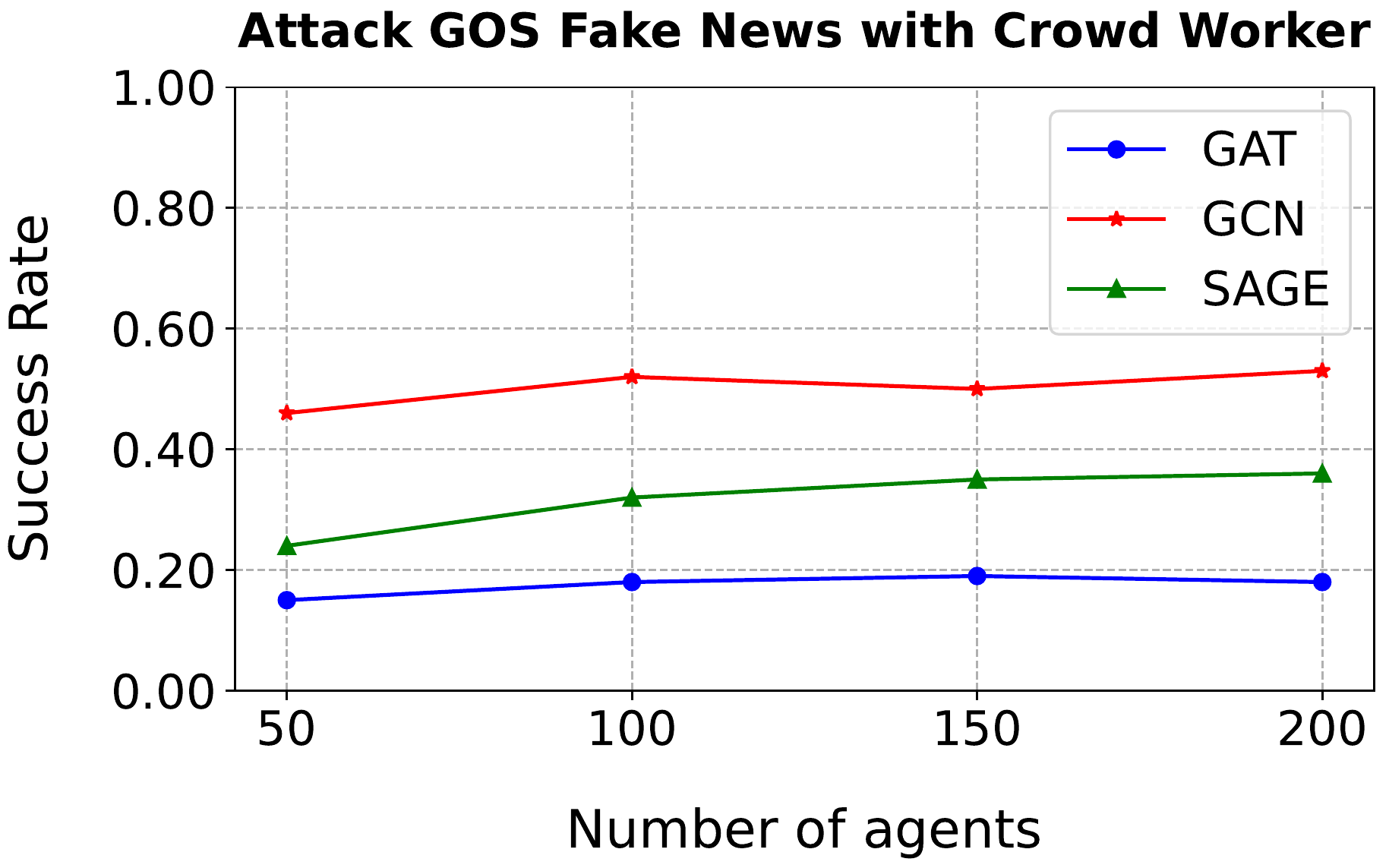}
    \end{subfigure}}
    \captionsetup{width=0.95\linewidth}
    \vspace{0.1cm}
    \caption{\textit{Indirect} attack performance of different types of agents on fake news in Politifact and Gossipcop datasets. Performance on GAT, GCN, and GraphSAGE are marked in blue, red, and green respectively.}
    \Description[Attack Agent Ablation Study]{Attack performance of different types of agents on fake news in Politifact and Gossipcop datasets. Performance on GAT, GCN, and GraphSAGE are marked in blue, red, and green respectively.}
    \vspace{-0.10cm}
  \label{fig:agents}
\end{figure*}

\subsection{RQ1: Performance of MARL}
\label{subsec:performance}
Since attackers are more likely to use indirect attacks than direct attacks to evade detection in practice, we study \textbf{targeted indirect attacks} on both fake and real news in Politifact and Gossipcop.
Table ~\ref{tab:performance} reports the attack performance of MARL compared to the baselines. 
From the table, we make the following observation:
\begin{itemize}[leftmargin=*]
    \item We can see that MARL improves the overall attack performance on fake news across both datasets. Compared to the Random-Edge baseline, MARL improves the success rate of targeted fake news attacks by [57.6\%, 51.1\%, 53.6\%] respectively for [GAT, GCN, GraphSAGE] on Politifact, and [250\%, 128.6\%, 44\%] on Gossipcop.
    \item We find that \textbf{news posts close to the decision boundary are easier to attack}. For example, when we compare the success rate of indirect attacks on fake news that falls into the decision boundary of [0.5, 0.7] to those outside of the decision boundary, we see an increase in success rate from 0.18 to 0.33 when attacking fake news in Politifact dataset on GAT detectors.
    \item Another interesting finding is that \textbf{GCN is more sensitive to edge perturbations compared to GAT and GraphSAGE}. Attackers can achieve fairly good performance on GCN with only a small amount of edges added to the graph. For instance, we can reach a success rate of 0.48 with just 210 edges added when attacking fake news on Politifact.
    Comparably, with the same amount of attack budget, the success rate on GAT and GraphSAGE detectors are 0.12 and 0.14 respectively, which are much lower than GCN. Previous works \cite{dai2018adversarial, chen2021understanding} have also shown that GCNs are vulnerable to structural adversarial attack due to the low breakdown point of their weighted mean aggregation method.
\end{itemize}

\subsection{RQ2: Attack Performance Analysis}
In this subsection, we answer RQ2 and provide insights on the factors that affect MARL's attack performance. Specifically, we provide analysis based on agent types and news types.

\subsubsection{Agent Types}
Figure ~\ref{fig:agents} shows the ablation study results on single agent RL. Specifically, we use RL-A1, RL-A2, and RL-A3 methods to carry out targeted indirect attacks on fake news in Politifact and Gossipcop datasets with increasing attack budgets by using more agents. We make the following observations:

\begin{itemize}[leftmargin=*]
    \item The overall attack performance increases with incremental attack budget for all three types of agents.
    \item The performance gain slows down after hitting a threshold. Therefore, attackers need to select the optimum number of agents to perform indirect attacks.
    \item Crowd worker agents achieve better performance than bot and cyborg agents on all three GNNs given the same amount of attack budgets. This is expected since crowd worker agents have stronger influence and their social posts are connected to real news. Therefore, they exert more influence on fake news.
\end{itemize}

Based on the above observation, we divide user nodes into ``good'' and ``bad'' groups. Specifically, we put users who have more than 80\% of the news they shared being fake into the ``bad'' group and users who have less than 20\% of the news they shared being fake into the ``good'' group. 

\begin{table}[b!]
\vspace{-0.5cm}
  \centering
  \captionsetup{width=0.90\linewidth}
  \caption{\textit{Indirect} targeted attack on fake news in Politifact and Gossipcop based on their news degrees using crowd worker agents.}
  \vspace{0.1cm}
  \resizebox{0.90\linewidth}{!}{%
  \begin{tabular}{@{}ccccccc@{}}
    \toprule
    \multirow{2}{*}{\textbf{News Degrees}}                                       & \multicolumn{3}{c}{\textbf{Politifact}} & \multicolumn{3}{c}{\textbf{Gossipcop}} \\ \cmidrule(l){2-7}
                                               & GAT  & GCN  & SAGE & GAT  & GCN  & SAGE \\ \midrule
    \textbf{[0, 10)}                    & 0.16 & 0.30 & 0.21 & 0.14 & 0.22 & 0.25 \\
    \textbf{[10, 100)} & 0.14 & 0.15 & 0.11 & 0.11 & 0.13 & 0.12 \\
    \textbf{[100, $\infty$)}                 & 0.03 & 0.06 & 0.03 & 0.02 & 0.02 & 0.05 \\ \bottomrule
  \end{tabular}}
  \label{tab:degree}
\end{table}

\subsubsection{News Types}
Intuitively, we conjecture that the news post node with a higher degree is more robust to attacks than those with lower degrees. 
To verify this hypothesis, we attack different groups of fake news in Politifact and Gossipcop according to their node degree. For this experiment, we use 10 crowd worker agents for Politifact and 50 crowd worker agents for Gossipcop respectively.
As shown in Table~\ref{tab:degree}, it is significantly harder to attack news with a higher degree across all three GNNs. 
Even on the most vulnerable GNN (i.e. GCN), MARL has a significant performance decrease (80\% on Politifact and 90.9\% on Gossipcop) when attacking fake news with a degree of less than 10 compared to the news with a degree more than 100.
Another observation is that GAT is more robust than GCN and GraphSAGE on news with degree between 10 and 100. As shown in Table \ref{tab:degree}, GAT only has a performance drop of 12.5\% and 21.4\% on Politifact and Gossipcop respectively when increase news degree from less than 10 to less than 100. Whereas the performance drop of GCN and GraphSAGE are almost halved on both datasets. This is likely due to the attention mechanism of GAT making it less sensible to degree changes.

\begin{figure}[!tbp]
  \centering
\vspace{0.1cm}
  \resizebox{0.88\linewidth}{!}{%
  \begin{subfigure}[b]{0.23\textwidth}
    \centering
    \includegraphics[width=\textwidth]{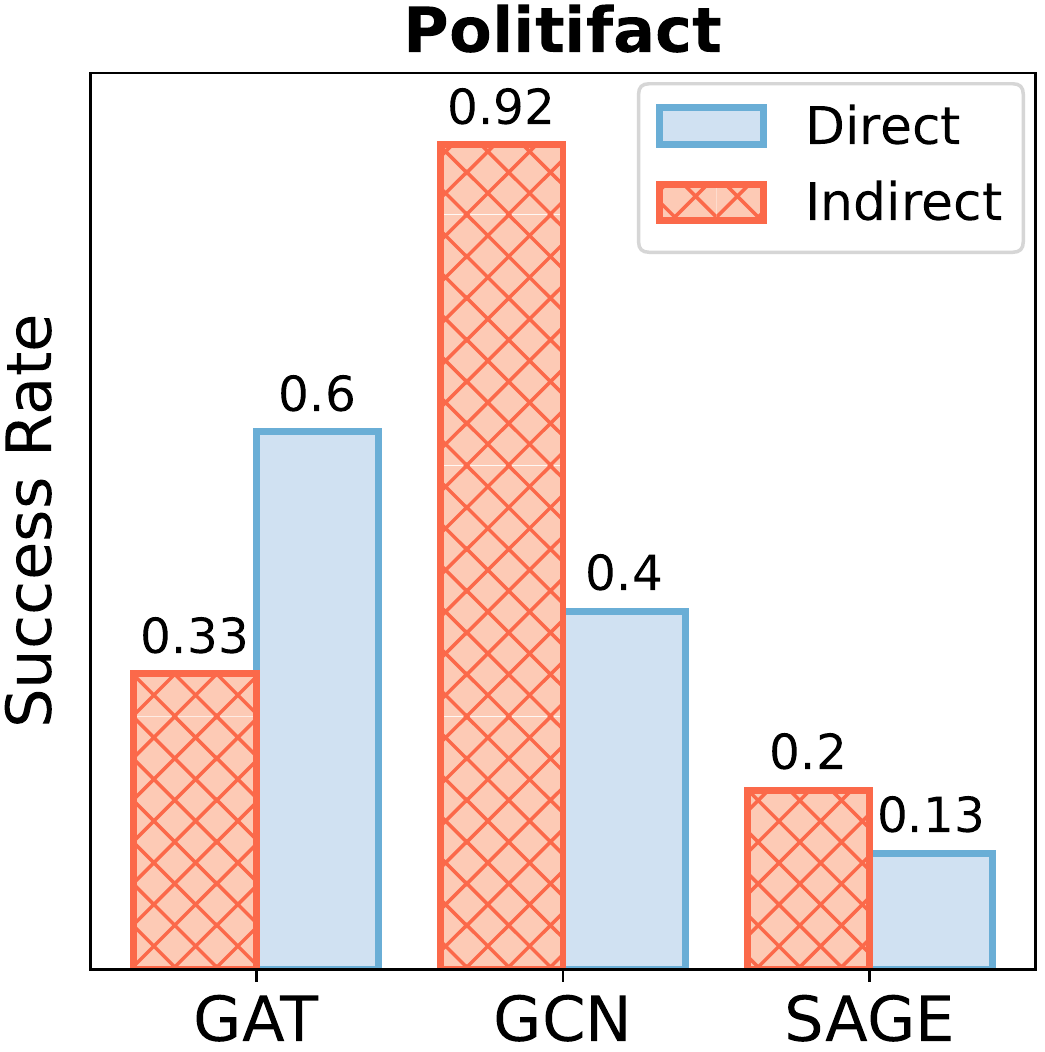}
    \label{fig:pol_compare}
  \end{subfigure}
  \hspace{0.25cm}
  \begin{subfigure}[b]{0.23\textwidth}
    \centering
    \includegraphics[width=\textwidth]{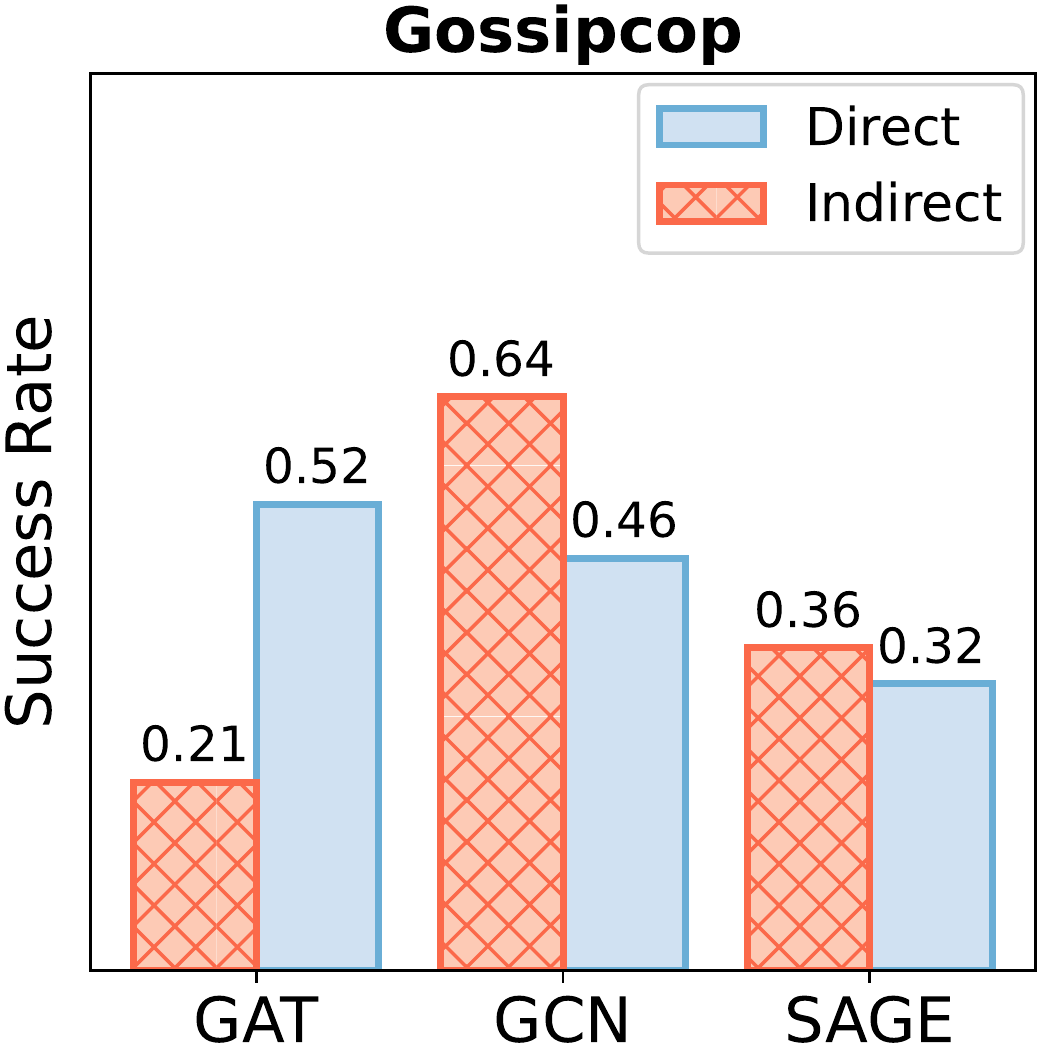}
    \label{fig:gos_compare}
  \end{subfigure}}
  \caption{Comparison between the \textit{direct} and \textit{indirect} attack on Politifact and Gossipcop on fake news with degrees less than 10 using ``good'' users.}
  \Description[Direct vs. Indirect Attack]{Comparison between the direct and indirect attack on Politifact and Gossipcop on fake news with degrees less than 10 using ``good'' users.}
  \label{fig:compare}
  \vspace{-0.5cm}
\end{figure}

\subsection{RQ3: Direct vs. Indirect}
Recall from Section~\ref{sec3:capability}, direct attacks mean that attackers modify the edges directly linked from user nodes to target news nodes.
Although direct attacks are more obvious in real-world scenarios, they can achieve better performance than indirect attacks due to the direct perturbation of graph structure.
For this experiment, we sampled 25 good users for Politifact and 250 good users for Gossipcop to perform targeted attacks on fake news in both datasets.
Figure ~\ref{fig:compare} shows the comparison between direct and indirect attacks. 
We can see that direct attack improves the performance on GAT by 81.8\% and 147.6\% on both datasets respectively. 
However, we see a decrease in performance on GCN and GraphSAGE detectors across both datasets. Especially on GCN detectors in the Politifact dataset.

\begin{figure}[!tbp]
  \centering
  \begin{subfigure}[b]{\textwidth}
    \includegraphics[width=0.23\textwidth]{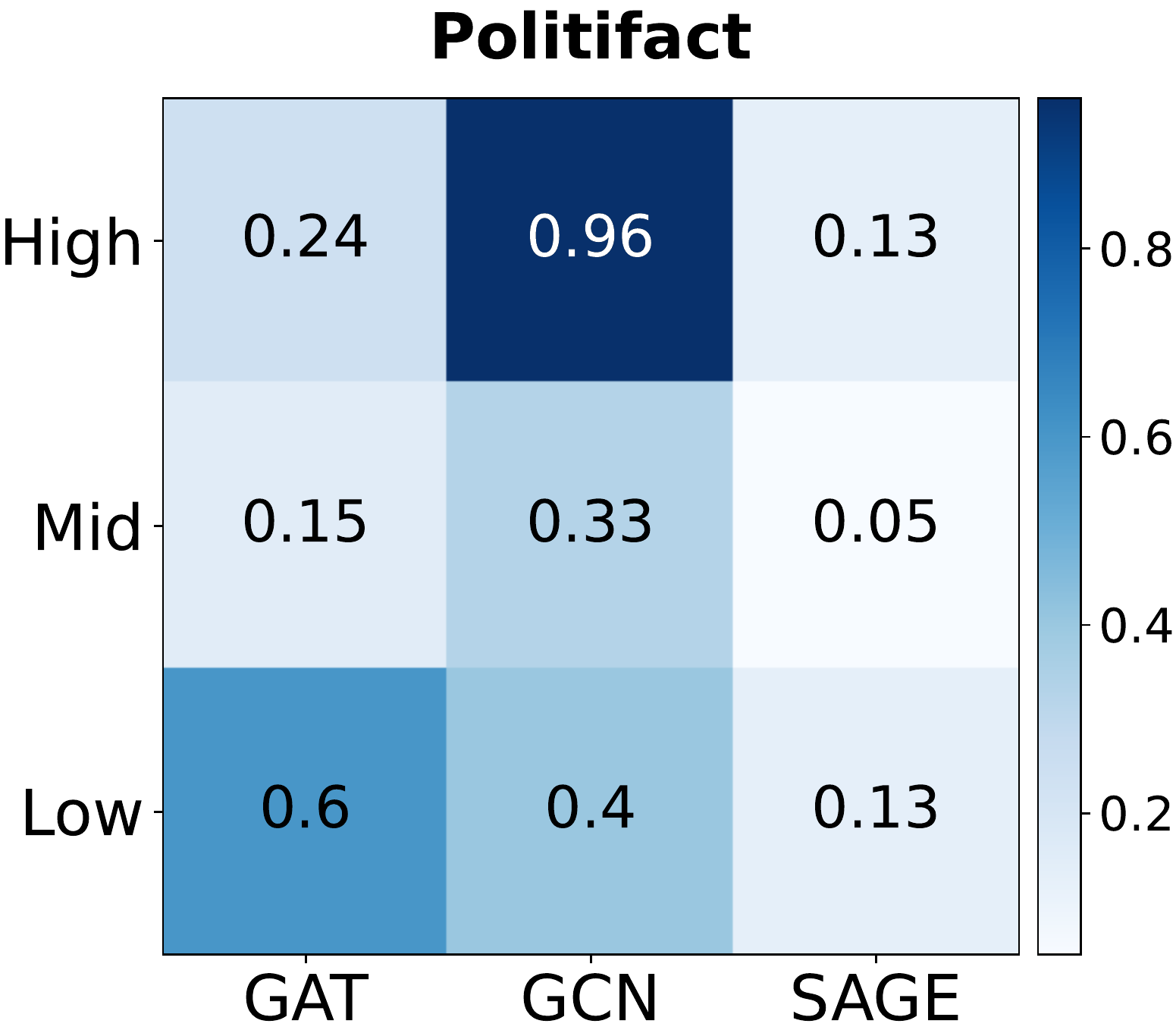}
    \includegraphics[width=0.23\textwidth]{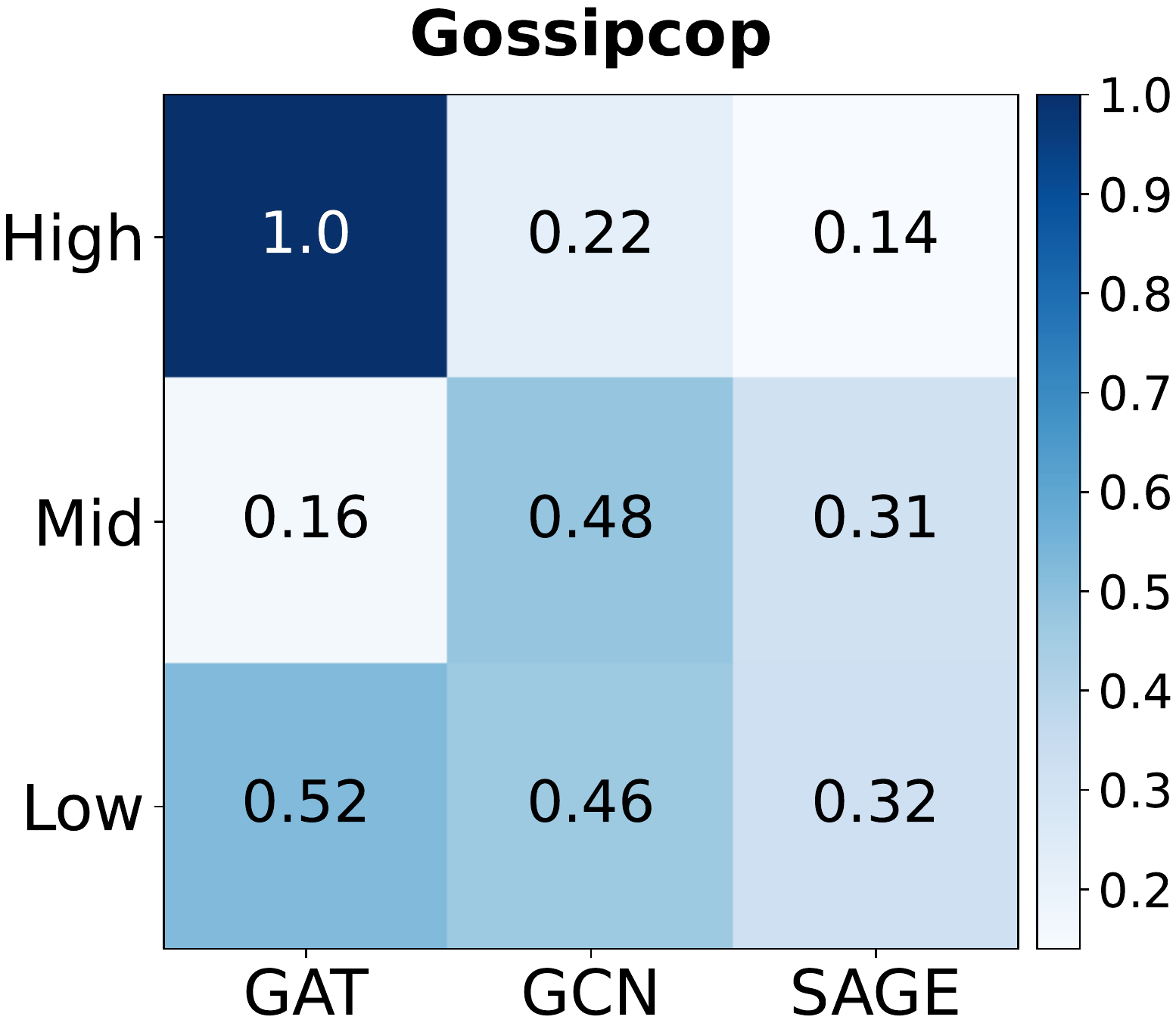}
  \end{subfigure}
  \vspace{0.05cm}
  \captionsetup{width=0.98\linewidth}
  \caption{Results of \textit{direct} attack across different types of news based on their degrees. A brighter color suggests better attack performance.}
  \Description[Direct attack on fake news with different news degrees]{Results of direct attack across different types of news based on their degrees. A brighter color suggests better attack performance.}
  \label{fig:heatmap}
  \vspace{-0.8cm}
\end{figure}

Based on the observation from Figure \ref{fig:compare}, we are interested in whether the performance of direct attack behaves similarly across news with different degrees. For this experiment, we categorize fake news in Politifact and Gossipcop based on their news degrees: news with less than 10 tweets is classified as low; news with more than 100 tweets is classified as high; and news in between as mid. We use the same agent configuration as in Figure \ref{fig:compare} to attack fake news in both datasets.
Figure ~\ref{fig:heatmap} shows that direct attack is effective on news with low degrees on GAT detectors, while it is less effective on GraphSAGE detectors regardless of the news degrees.

\subsection{RQ4: Countermeasures against Attacks}
Based on our experiment findings, we discuss the countermeasures for fraudsters that manipulate news social engagement from two perspectives.
\textbf{1)} From the machine learning security perspective, there are fruitful research works on defending against graph adversarial attacks~\cite{sun2018adversarial}.
Approaches like adversarial training~\cite{feng2019graph}, anomaly detection~\cite{duan2020aane}, and robust GNN models~\cite{geisler2020reliable, jin2020graph} can be leveraged to defend the attacks. 
\textbf{2)} From the practical perspective, social media platforms should pay equal attention to both ``bot'' and seemingly ``good'' users. As shown in the experiments, attackers can leverage users' good posting history to carry out a successful targeted attack on fake news to foul GNN-based detectors.
Since the indirect attack is effective against many GNN detectors, this suggests that the platform should monitor more engagement activities of accounts engaged with the target news instead of the target news itself.
Experiment results also show that there is no universally robust model which prompts the platform to adopt diverse trust and safety models.

\section{Conclusion and future work}
\label{sec:conlusion}
In this paper, we aim to understand the vulnerability of graph neural network-based fake news detectors under structural adversarial attacks. 
To the best of our knowledge, this is the first work to attack GNN-based fake news detectors. 
This paper aims to provide insights on how to develop a more robust GNN-based fake news detector against adversarial attacks in the future.
We leveraged a multi-agent reinforcement learning framework to mimic the attack behavior of fraudsters in real-world misinformation campaigns. Our experiment results show that MARL improves overall attack performance compared to our baselines and is highly effective against GCN-based detectors.

Even though we have some promising results from the experiments, this paper has two major limitations: 
\textbf{1)} This work only employs a simple heuristic to select users for action aggregation.
\textbf{2)} The search space of the Q network is considerably large and results in a high computational cost on larger datasets like Gossipcop.
Therefore, there are several interesting directions that need further investigation. The first one is to automate the process of selecting optimal agents for action aggregation.
The second one is to reduce the deep Q network's search space effectively. 
Finally, we used a vanilla MARL framework in this paper. It would be interesting to explore a more complex MARL framework for this task. 

\section{Ethical Statement}
The Twitter data used in this paper are obtained from Twitter API and meet Twitter user agreement.
Although we proposed an adversarial attack framework against GNN-based fake news detectors, our intention is to probe and enhance the robustness of existing detectors.
Therefore, we do not endorse this work to be used for unethical purposes in any shape or form.

\begin{acks}
This work is supported in part by NSF III-1763325, III-1909323, SaTC-1930941, III-2106758, SaTC-2241068, and a Cisco Research Award.
This work was done before Yingtong Dou joined Visa Research.
\end{acks}

\balance
\bibliographystyle{ACM-Reference-Format}
\bibliography{main}


\end{document}